\documentclass[conference]{IEEEtran}
\IEEEoverridecommandlockouts
\usepackage{cite}
\usepackage{amsmath,amssymb,amsfonts}
\usepackage{color}
\usepackage{algorithmic}
\usepackage{graphicx}
\usepackage{textcomp}
\usepackage{url}
\usepackage{xcolor}
\def\BibTeX{{\rm B\kern-.05em{\sc i\kern-.025em b}\kern-.08em
    T\kern-.1667em\lower.7ex\hbox{E}\kern-.125emX}}
\begin{document}

\title{A Dynamic and Cooperative Tracking \\System for Crowdfunding
}

\author{Kai Zhang$^{\dag}$,  \hspace{1mm}Hongke Zhao$^{\ddag}$,  \hspace{1mm}Qi Liu$^{\dag}$, \hspace{1mm}Zhen Pan$^{\dag}$, \hspace{1mm}Enhong Chen$^{\dag}$\\
		$^\dag$University of Science and Technology of China \\
		$^\ddag$The College of Management and Economics, Tianjin University \\
		 \{sa517494, pzhen\}@mail.ustc.edu.cn, hanqinghongke@gmail.com, \{qiliuql, cheneh\}@ustc.edu.cn
}

\maketitle

\begin{abstract}
Crowdfunding is an emerging finance platform for creators to fund their efforts by soliciting relatively small contributions from a large number of individuals using the Internet. Due to the unique rules, a campaign succeeds in trading only when it collects adequate funds in a given time. To prevent creators and backers from wasting time and efforts on failing campaigns, dynamically estimating the success probability of a campaign is very important. However, existing crowdfunding systems neither have the mechanism of dynamic predictive tracking, nor provide the real-time campaign status for creators and backerspre on the platform. To address these issues, we develop a novel system, which contains a dynamic data-driven approach to tracking the success probability and status. We demonstrate the following scenarios using our system. First, users can utilize our system to analyze the emotion of incremental reviews so as to understand backers' perspectives of the campaign in time. Meanwhile, our system visualizes the statistic number of positive and negative reviews. On this basis, our system can dynamically track the success probability of each campaign.
\end{abstract}

\section{Introduction}
Crowdfunding is the practice of funding a campaign or venture by raising monetary contributions from a large number of people. Recent years have witnessed the rapid development of crowdfunding platforms, such as Indiegogo.com. Despite the huge success of crowdfunding platforms, the percentage of campaigns that succeed in achieving their desired goal amount is only around $40\%$. According to the ``all or nothing" policy, once a campaign fails, the creator will lose all the funds raised before, and all efforts made on this campaign will turn into nothing \cite{b1}. Meanwhile, if the investment campaign fails, the backers who participated in the campaign can recover the funds but may waste lots of time and opportunity cost. Fortunately, dynamic tracking of the campaign financing process can help creator better adjust campaign design timely and accurately, which can potentially lead to a better survival environment for campaigns and provide more benefits information for creators and backers. 
\begin{figure} [t]
	\centering
	\includegraphics[scale=0.45]{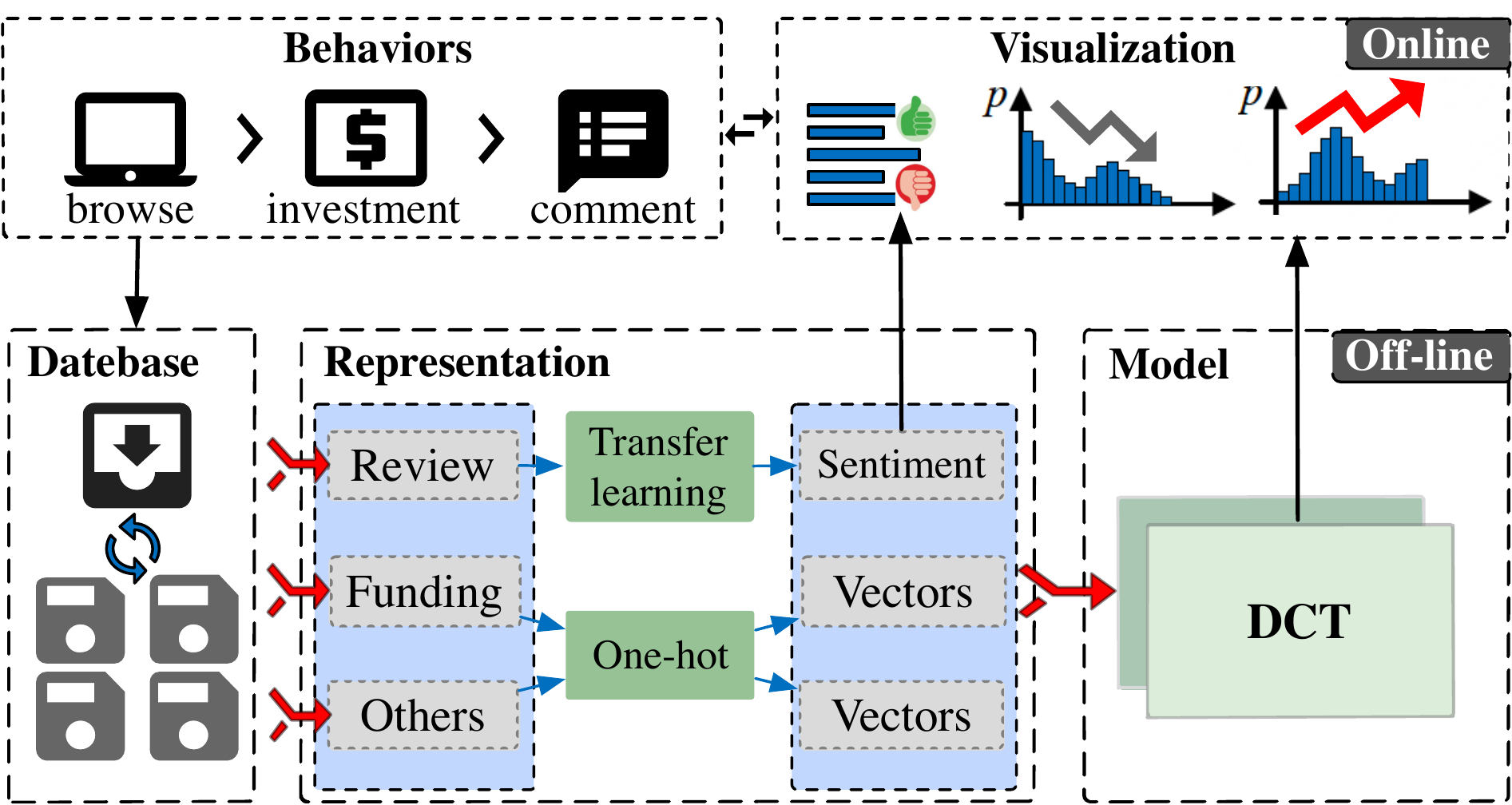}
	\caption{DCT System Flowchart.}
	\label{fig:frame}
\end{figure}

As it is crucial to track the dynamics of the campaign, much research attention has been attracted from both academia and industry. In the literature, many methods have been proposed to solve the problem of predicting the campaign state, especially various solutions for multi-feature learning have been designed\cite{b2}\cite{b3}. Despite the huge success of the existing methods, there are still limitations. Firstly, most of the previous systems focus on predicting the final success rate of campaigns and ignoring the dynamic process, such as the amount of money received each day and the sentiment tendency of the reviews which can express backer's preference more directly. Thus, we fully mine these two kinds of information in our system to track and predict the daily success rate of the campaign during the fundraising process. Secondly, most of the existing methods merely rely on historical and static information rather than considering the incremental information which can better model the process of campaigns. Therefore, we argue to model the daily incremental information into our system to achieve dynamic tracking of the campaign.

Through the above analyses, we design and implement a novel system named \textbf{D}ynamic and \textbf{C}ooperative \textbf{T}racking (DCT) System\footnote{\scriptsize A brief demonstration of the DCT system, \url{https://youtu.be/ZV9kWKkX7Z8}} which is combined with static and dynamic information 
cooperatively to track the incremental states of campaigns. Specifically, we apply a transfer learning model to do sentiment classification on the campaign reviews and show the statistic of emotions. Furthermore, we present all the tracking results of the campaign to users dynamically and interactively through the online visualization part. In summary, our system encompasses the following benefits:
\begin{itemize}
\item It analyzes the emotion of each review and shows the statistics of review emotion.
\item It implements dynamic tracking and visualization of the campaign that is not available in previous systems.
\item It integrates a novel tracking model by the combination of historical static and dynamic incremental information, which performs better than the conventional methods.
\end{itemize}

\begin{figure} [t]
	\centering
	\includegraphics[scale=0.2]{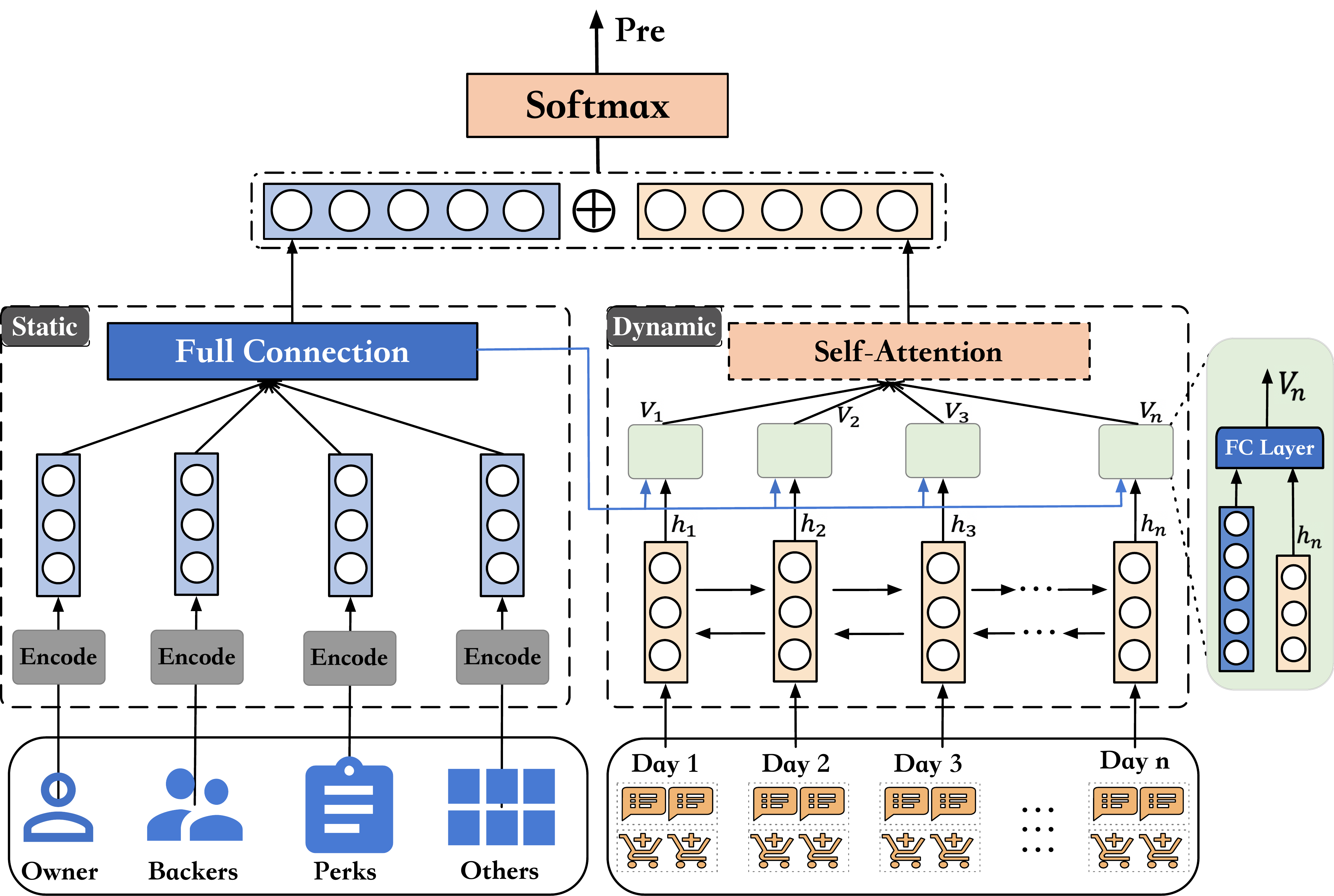}
	\caption{System technical part, DCT model.}
	\label{fig:model}
\end{figure}
\section{System Overview}
In this section, we present the architecture of DCT System which is illustrated in Fig.~\ref{fig:frame}. DCT System provides functionalities for dynamic data-driven analysis, which we will detail below. The entire system overview can be divided into two parts\cite{b4}: offline part is designed for training our model to provide prediction results for visualization of the online part, and the online part responds to user's behaviors and then makes corresponding dynamic information demonstrations.

\subsection{Offline Part}
\subsubsection{Data Collection} 
We collect massive real-world crowdfunding data with detailed daily transaction information from Indiegogo. After carefully crawling the crowdfunding dataset, we get 12,328 campaigns with almost 582,560 reviews and 20 other kinds of attributes (e.g., \emph{goal, duration and information of owner}) and then store them in a local database.

\subsubsection{Data Preprocess} 
In order to solve the problems mentioned above, we do some necessary preprocessing operations on the dataset. First, we extract the daily review information and funding information of the campaign since we believe that these incremental data can be better utilized\cite{b5}. Specifically, for reviews, in order to solve the problem of missing sentiment labels, we extract all reviews from other attributes of the campaign and use the existing transfer learning method\cite{b6} to do sentiment classification for them. Besides, for funding data, we will pre-process them based on the amount of money which campaign received each day and then map these data into feature vector representations with the one-hot method. Similarly, we use the same mapping method to process the other attribute information. After obtaining the review and funding feature representations, we are going to dynamically model them in the next subsection.

\subsubsection{Model Design} 
In this subsection, we first detail how to model the campaign information and then present how the model is trained according to the strategies we proposed. As stated in the previous section, our model mainly mines dynamic review information and funding information, and also combines other static information of the campaign at the same time. Thus, we propose a \textbf{D}ynamic and  \textbf{C}ooperative \textbf{T}racking (DCT) model to collaboratively model the static and dynamic features, i.e., fundings and reviews. Fig.~\ref{fig:model} shows the detailed training process of our model and we will give the details in section III.

\subsection{Online Part}
The online part in Fig.~\ref{fig:frame} is designed as a web so that users can play with the system in real cases. Our system is actually a crowdfunding platform, which is first designed and implemented as a website and then given a visualized dynamic prediction presentation of the campaign. As we can see the visualization frame from Fig.~\ref{fig:frame}, there are three visual pictures, which the first one is denoted as review sentiment visualization and the next two are campaign's dynamic prediction visualization. According to the online part of the visualization, users can better understand the current state of the campaign.

\section{Technical Background}
In this section, we present the technical details on how to implement a dynamic tracking system. Formally, we first present the techniques of historical static data modeling followed by an overview of the dynamic incremental data modeling. Then the technical details of the dynamic tracking process are introduced.

\subsection{Static Modeling} \label{AA}
As mentioned above, each campaign on the crowdfunding platform contains approximately twenty static properties, which can roughly be divided into four main categories (i.e., \emph{owner, backer, perks and other attribute information}). As the static part shown in Fig.~\ref{fig:model}, we encode these four category features as the vector representation separately and normalize them to eliminate the dimensional influence among feature data. Finally, we combine these four feature vectors into a global feature vector which we denote as $S_r$. Note that after the \emph{full-connection} layer, the static feature representations of the campaign are fixed, which means that they do not change along the campaign progresses. Thus, we also regard this part of information as historical static data.

\subsection{Dynamic Modeling}  
In this subsection, we show how to model dynamic data (i.e., \emph{the funding data and the review data}). As mentioned in the above section, we first need to tag the review information of the campaign with sentiment label through transfer learning method. Here, we choose the pre-trained model, which is trained base on the Amazon review dataset for cross-domain sentiment classification, to do sentiment analysis on our campaign reviews. After getting the sentiment representation, we summarize the reviews in each day so that we can analyze the history and process information of campaigns by combining them. Similarly, we use the same summary method to process the fundings and map them to feature vectors by the one-hot method. Finally, we connect both daily review feature representation and funding feature representation as the daily dynamic features of the campaign.

Owing to our daily dynamic features have strong time series characteristics, we adopt a deep learning method named Long-Short Term Memory networks (\emph{LSTM})\cite{b7} to learn hidden states of each day's feature, because it performs well in learning long-term dependencies and can effectively solve gradient vanishing and expansion problems. Specifically, given the daily feature vector representation (e.g., $r_1, r_2, r_3, ..., r_n$)  of campaign as the input, \emph{LSTM} updates the cell vector sequence $c_t$ and the hidden state $h_t$ from $t = 1$ to $n$. After the initialization, at \emph{t-th} interaction step, the hidden state $h_t$ of each interaction is updated by the previous hidden state $h_{t-1}$ and the current day's feature vector $r_t$ as:
\begin{equation}
\small
	\begin{split}
	&i_t\ =\ \delta(W_{ei}{r_t}\ +\ W_{hi}h_{t-1}\ +\ {\widehat{b}_i})\ ,\\
	&f_t\ =\ \delta(W_{ef}{r_t}\ +\ W_{hf}h_{t-1}\ +\ {\widehat{b}_f})\ ,\\
	&c_t\ =\ f_t\cdot c_{t-1}\ +\ i_t\cdot \tau(W_{ec}{r_t}\ +\ W_{hc}h_{t-1}\ +\ {\widehat{b}_c})\ ,\\
	&o_t\ =\ \delta(W_{eo}{r_t}\ +\ W_{ho}h_{t-1}\ +\ {\widehat{b}_o})\ ,\\	
	&h_t\ =\ o_t\ \cdot\ {tanh}(c_t)\ ,
	\end{split}
	\label{eq:lstm}
\end{equation}
where $i_t,f_t$ and $o_t$ are the input, forget and output gates at \emph{t-th} step respectively. $r_t$ is the feature embedding vector. $c_t$ is the cell memory and $h_t$ is the output. $\delta(\cdot)$ is non-linear activation function which is stated as $sigmoid$ in this paper. Dot $\cdot$ denotes the element-wise multiplication between vectors. $W_\ast$ denotes weight matrices, $\widehat{b}_\ast$ is the bias vectors.

\subsection{Combination \& Prediction} 
After the \emph{LSTM} layer, the daily reviews and funds representation is transformed to the \emph{hidden states} (i.e., $h_s=\{h_1, h_2, h_3, ..., h_n\}$). In order to model the static features and dynamic features more efficiently and cooperatively, we connect the static feature vector (i.e., $S_r$) with each hidden state of dynamic features, and state the result feature vectors as $V_s=\{V_1, V_2, V_3, ..., V_n\}$. As discussed before, the incremental data of each day has a different influence on the prediction. For example, if the majority of evaluation on the 10th day is negative, the influence of the day may be greater for the final prediction since people are more susceptible to negative emotions than positive emotions. Fortunately, attention mechanisms can highlight different parts of the input by assigning weights to encoding vectors in each step of feature representation. Thus, we generate the attention vector $\alpha_i$ as: 
\begin{equation}  
	\begin{split}
	&\alpha_i\ =\ \frac{exp(\ {tanh}({V_i} \cdot\ W_s\ +\ {\widehat{b}_s}))}{\sum_{i=1}^{n}exp(\ {tanh}({V_i} \cdot\ W_s\ +\ {\widehat{b}_s}))}\ ,
	\end{split}
	\label{eq:weight}
\end{equation}
where $W_s$ and $\widehat{b}_s$ are weight matrix and bias matrix respectively. $tanh$ is a non-linear function. After computing the attention weights of the daily feature, we can get the final representation of dynamic features. The formula expression is:
\begin{equation}
	\begin{split}
	&D_r\ =\ \sum_{i=1}^{n}{\alpha_i}{V_i}\ .
	\end{split}
	\label{eq:scorefunction}
\end{equation}

At last, we combine the historical static data representation $S_r$ with the final dynamic incremental data representation $D_r$ and process them through \emph{softmax} layer to do dynamic prediction of campaigns.

\section{Demonstration Details}
The demonstration of our system shows three parts of the real fundraising process situation for campaign analyze, including emotional classification, sentiment statistics, and dynamic tracking. The system is implemented in Python and PHP on a MySQL database running on a cloud computing platform. DCT System is built based on a real Indiegogo dataset and the demonstration attendees can interact with our system to get more dimensional campaign information. 

\begin{figure} [t]
	\centering
	\includegraphics[scale=0.3817]{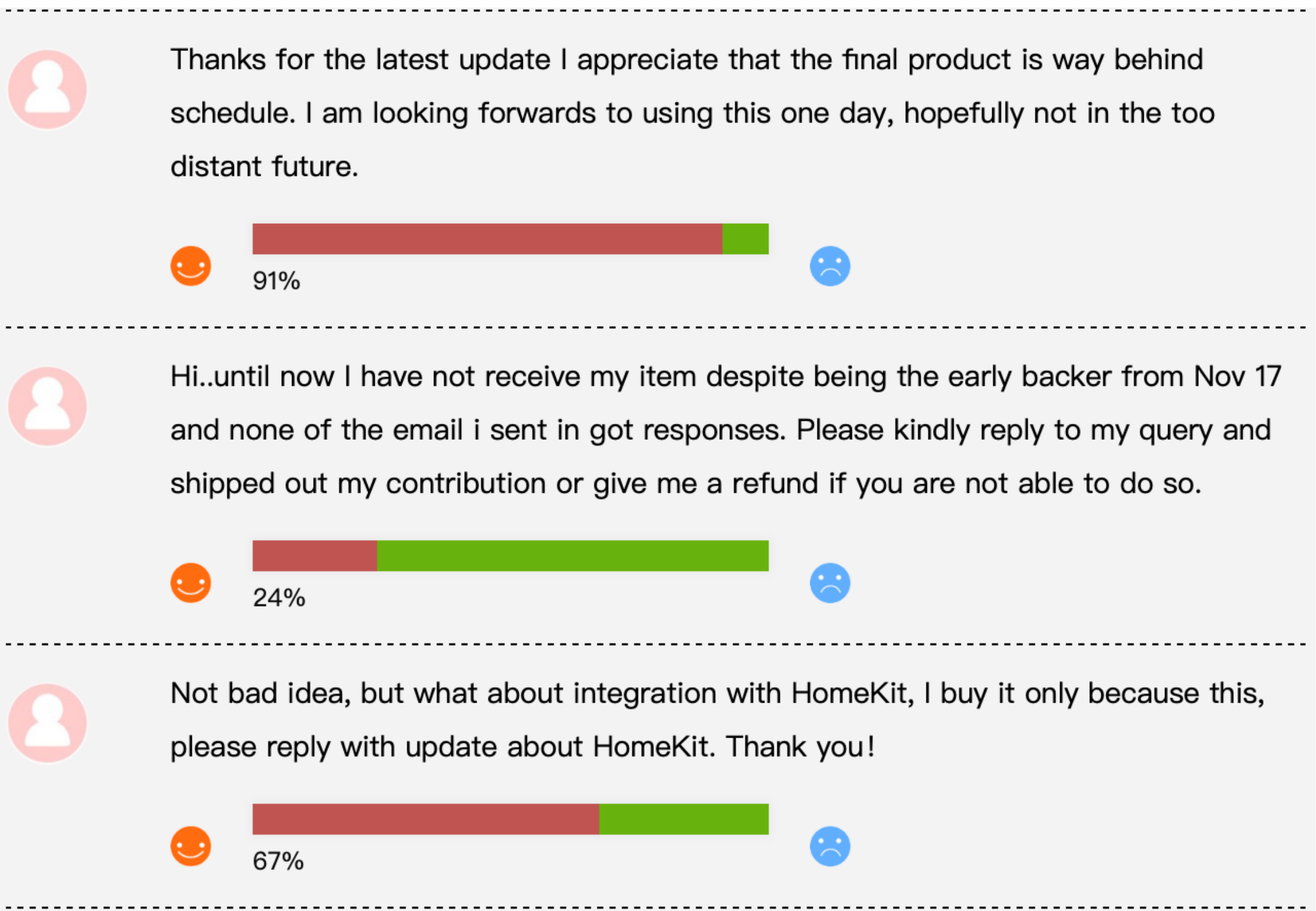}
	\caption{Sentiment classification with transfer learning method.}
	\label{fig:review}
\end{figure}

\subsection{Emotional Classification}\label{AA}
As mentioned above, we apply a transfer learning model to do sentiment classification because our review data has no labels. Then, we need to visualize the sentimental tendencies so that users can get a more intuitive view of other peoples' perspectives of the campaign. For example, as shown in Fig.~\ref{fig:review}, we predict that the first review has a $91\%$ probability of negative emotions, which is similar to our subjective judgment. The same is true for other reviews.

\begin{figure} [t]
	\centering
	\includegraphics[scale=0.37134]{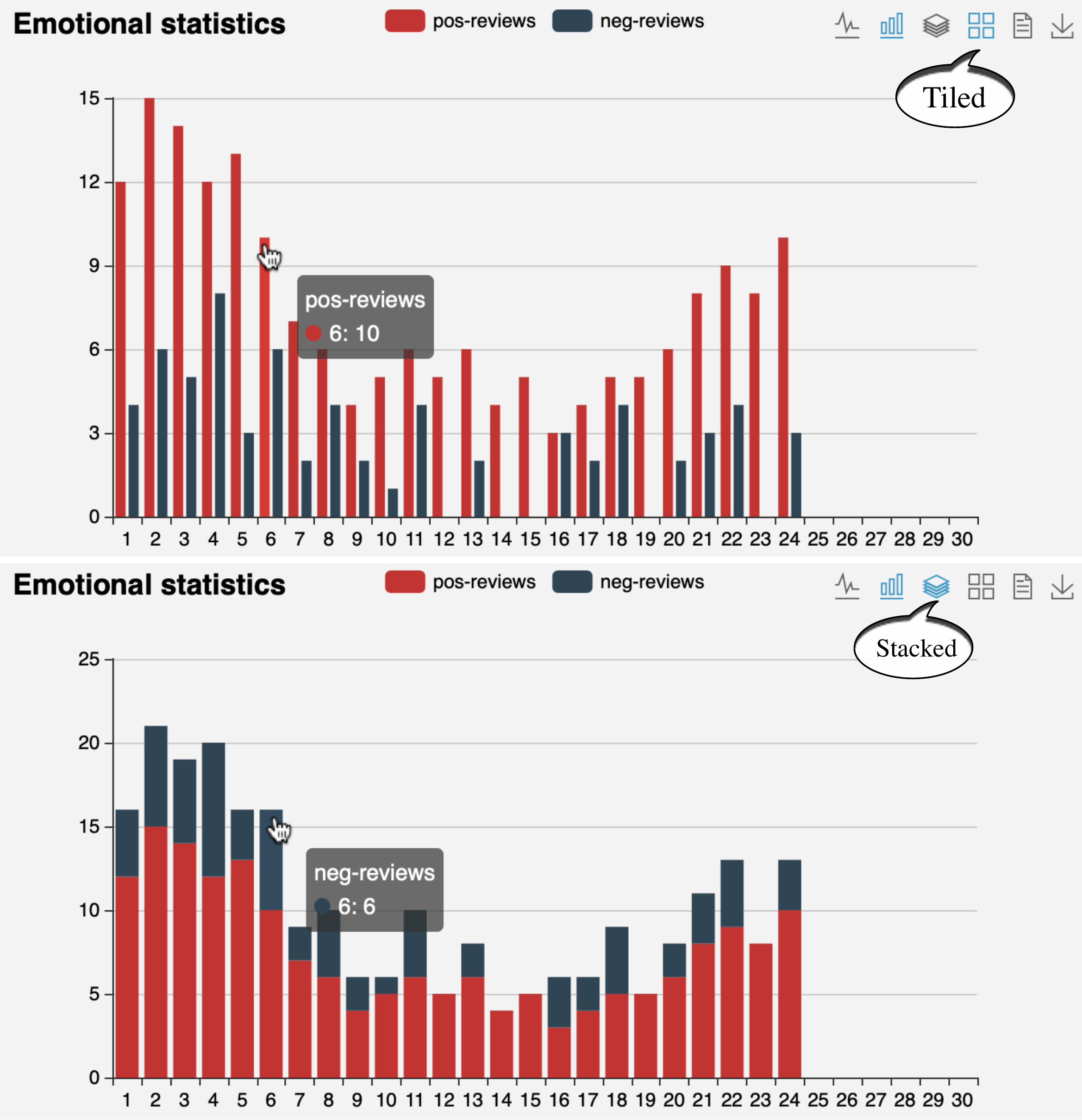}
	\caption{Sentiment analysis and statistical visualization.}
	\label{fig:statistic}
\end{figure}

\subsection{Sentiment Statistics}
After getting the sentiment label of the review, in order to track the emotional changes of backers, we perform the necessary review data analysis every day. And we present the results of the analysis in two patterns, i.e., \emph{tile pattern} and \emph{stack pattern}.
\subsubsection{Tile Pattern} 
As shown in the upper part of Fig.~\ref{fig:statistic}, we perform a statistical analysis of the predicted emotional sentiment and display it on the campaign review page. In this figure, the horizontal axis represents the number of days, and the vertical axis represents the number of reviews. Besides, the red histogram represents the number of positive reviews, and the gray represents negative numbers. Through this formal display, we can intuitively track the number of positive and negative reviews for the campaign every day. For example, we can see at a glance that the number of positive reviews received on the sixth day of the campaign is ten. 

\subsubsection{Stack Pattern} 
Stack pattern is another way to visualize which can more intuitively display the proportion of positive and negative emotions each day. From the lower part of Fig.~\ref{fig:statistic}, we can observe that six negative reviews are received on the sixth day of the campaign which is less than the number of positive reviews received. Due to space limitations, there are some other functions that we will show in the video.

\subsection{Dynamic Tracking}
Through the introduction of the above two sections, we can get the emotional information of the reviews and the number of positive and negative evaluations of the campaign each day. Then, we will further mine that information and combine them for the dynamic tracking. The details are shown in Fig.~\ref{fig:success} and expressed as follows.

\subsubsection{Postive\&Negative Emotion}  
As shown in this figure, the red histogram represents the positive emotional probability values. On the contrary, the gray histogram represents the probability that the item is a negative emotion. We only show results with a predicted probability of more than $50\%$ on that day, that is, if a red histogram is displayed on the day of the graph, the public opinion of the campaign on the day is generally positive and the gray is reversed. Note that this result is not obtained by simple statistics, but by model prediction. 

\subsubsection{Prediction with Funds} 
\begin{figure} [t]
	\centering
	\includegraphics[scale=0.34]{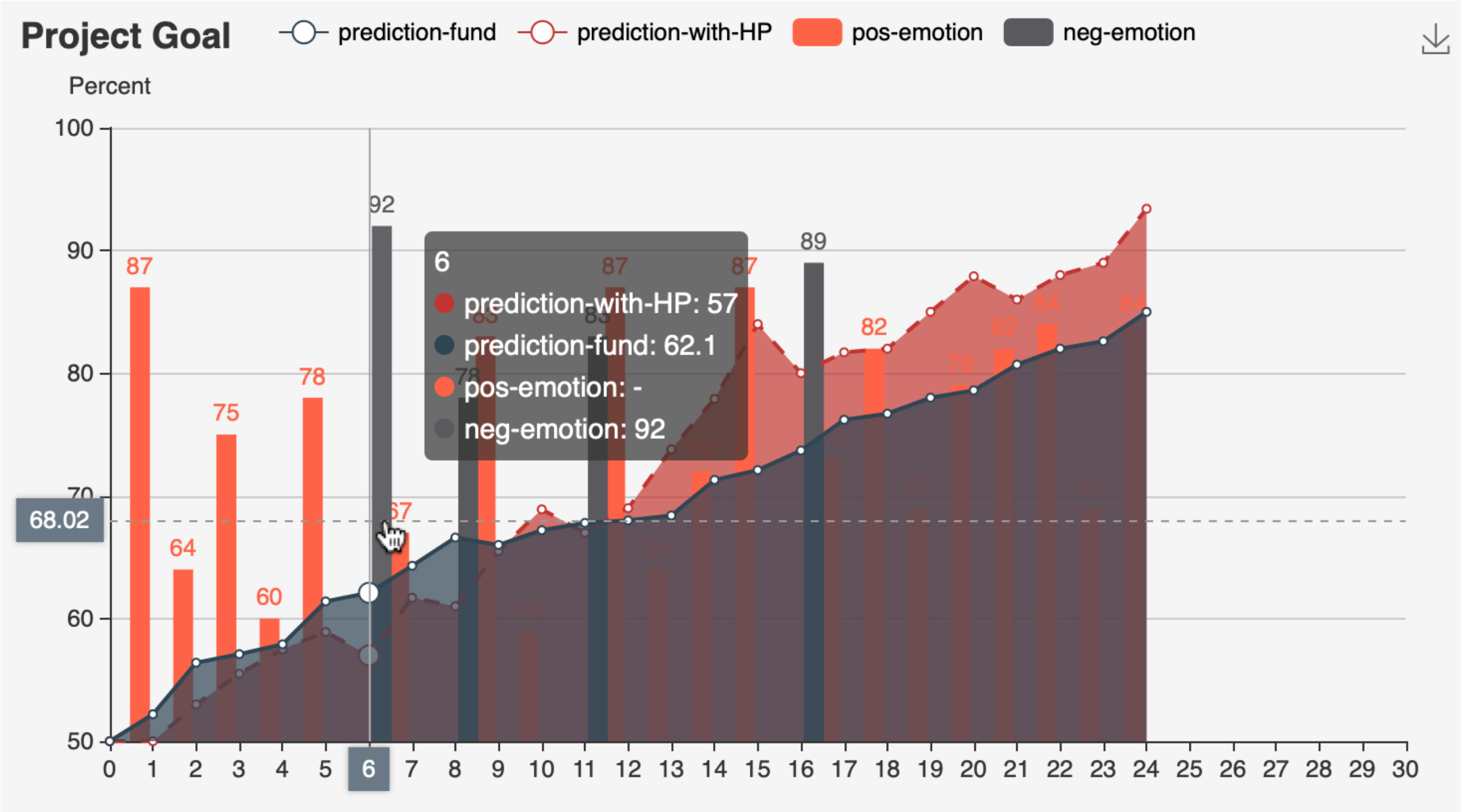}
	\caption{Dynamic tracking visualization.}
	\label{fig:success}
\end{figure}

In addition to the histogram, we also use a line chart to represent the results of campaign dynamic success prediction. The gray line chart represents the result of campaign predicting only using funds.  As we can see in Fig.~\ref{fig:success}, the trend of the gray line chart is always rising slowly and smoothly, which means that with the increase of funds, the success rate of the campaign is increasing.
  
\subsubsection{Prediction with Funds\&Reviews} 
At last, we use a red line chart to represent the success rate results of the campaign which predicted with funds and reviews. We can observe that the curve changes up and down in the overall trend but in some days when the overall reviews are negative, its trend is always falling, which means that reviews take an important role in the dynamic tracking. In addition, the method of adding review information to the final prediction probability is higher than the method of using only the funding data, which further illustrates the key role of real-time review information.
\section{Conclusion}
In this demonstration, we present a dynamic and cooperative tracking system by analyzing the user's sentiment tendency and investment preference. The system can dynamically give a daily success rate prediction based on the progress of the campaign, which significantly improves the backer's decision performance and provides a novel way to guide the process of campaigns for the creator. Further, we visualize all the dynamic predictions through our system, which means that users no longer need to carefully read the emotions of judging reviews, no longer need to manually count positive and negative reviews, everything can be easily and directly obtained from the visualization.

\end{document}